\documentclass[superscriptaddress,preprintnumbers,amsmath,amssymb,longbibliography,reprint,prb,citeautoscript,aps]{revtex4-1}
\setcitestyle{round}
\usepackage{graphicx}
\usepackage{dcolumn}
\usepackage{bm}
\usepackage{amsmath}
\usepackage{amssymb}
\usepackage{color}
\usepackage{hyperref}
\hypersetup{
    colorlinks=true,
    linkcolor=blue,
    filecolor=black,      
    urlcolor=blue,
    citecolor=blue,
}
\usepackage{braket}
\usepackage{soul}
\usepackage{relsize}
            
\usepackage{pifont}

\begin{document}

	\preprint{APS/123-QED}
	\title{Searching for various melting scenarios of 2D crystals}
	\author{Peng Hua}
	\author{Yilong Han}
	\email{yilong@ust.hk}
	\affiliation{Department of Physics, The Hong Kong University of Science and Technology, Clear Water Bay, Hong Kong, China}
	\date{\today}

    \begin{abstract} 
   In contrast to three-dimensional (3D) crystals that melt via a first-order transition, two-dimensional (2D) crystals can exhibit various melting scenarios under different temperatures, pressures, and particle interactions, particularly when they can form partially ordered intermediate phases between crystal and liquid. The existing 2D melting theories predict some scenarios, but do not exclude other possibilities. A recent simulation observed five melting scenarios including two new types by tuning the shape of polygonal particles. 

    \end{abstract}
    \maketitle

  The energy of long-wavelength waves in $d$-dimension scales as $L^d(2\pi/L)^2$ where $L$ is the length of the system. \cite{strandburg1988two} Therefore, thermal noises are not sufficient to induce long-wavelength distortions in large 3D systems, leading to the long-range order at low temperatures. The distortions are favorable in 1D, leading to the absence of an ordered structure. 2D is the critical dimension, and any infinitesimal thermal noise breaks the long-range translational order, which is essentially the Mermin-Wagner theorem. Thus, Peierls (1934) and Landau (1937) suggested that crystals cannot exist in 2D. Later Mermin (1968) denoted that the long-range orientational order can exist in 2D, and therefore 2D crystals can still exist. Intuitively, particles in low dimensions lack sufficient neighboring particles to correct the thermal-noise-induced deviations from their lattice sites, thus can only form less ordered structures. Note that the above arguments are for particles with short-range interactions. For long-range interactions, e.g., Coulomb force, particles have more effective neighbors, and thus the system is similar to be in a higher dimension.

    Low-dimensional crystals are soft with relatively more long-wave-length fluctuations. Hence, their melting behaviors can be distinct from 3D melting. The famous Kosterlitz-Thouless-Halperin-Nelson-Young (KTHNY) theory predicts a two-step continuous transition: crystal $\rightarrow$ hexatic phase $\rightarrow$ liquid. The hexatic phase is featured with quasi-long-range orientational order and short-range translational order induced by free dislocations.\cite{strandburg1988two} Chui and Saito showed that the KTHNY scenario occurs when the dislocation core energy is $>2.84~k_\textrm{B}T$ so that dislocations can disperse and induce a hexatic phase; When the core energy is $<2.84~k_\textrm{B}T$, dislocations condense into strings as grain boundaries and the melting is a one-step first-order transition.\cite{strandburg1988two,wang2016melting} Besides these two major theories based on topological defects (dislocations and disclinations), Glaser and Clark described 2D melting based on geometric defects.\cite{wang2016melting}  These theories do not exclude other possibilities such as those in Fig.~1A.
    
    The melting scenario depends on the particles' interaction and is difficult to predict in theory. In experiments, atomic and molecular monolayers are usually not freely suspended and strongly influenced by substrates.\cite{strandburg1988two} Large colloidal particles can avoid the substrate effects and serve as good model systems with measurable single-particle motions, but systematically tuning the particle's interaction is not easy. Thus, the understanding of particle-interaction effects on 2D melting is mainly obtained from systematic simulations.\cite{kapfer2015two,li2020attraction,lee2008effect,zu2016density,anderson2017shape,jiang2023five} Simulations show that repulsive disks with pair potential $U(r)\sim r^{-n}$ exhibit a continuous solid-hexatic transition and a first-order hexatic-liquid transition when $n \gtrsim 6$, including $n=\infty$ for hard disks.\cite{kapfer2015two} Such scenario for hard disks was subsequently confirmed in the colloid experiment. Soft particles with $n \lesssim 6$ exhibit the KTHNY melting scenario,\cite{kapfer2015two} following the experimental observations for colloids with $r^{-3}$ interaction.\cite{zahn1999two} Atoms and molecules have attractions which are also important to 2D melting. Simulations show that attractions promote a discontinuous melting transition with no hexatic phase and this effect vanishes at high temperatures since the thermal energy tends to overwhelm the attraction.\cite{li2020attraction} The one-step grain-boundary-mediated first-order 2D melting without substrate effects has not been observed experimentally until recently \cite{li2016modes} using attractive colloids. In addition, simulations show that long attraction range promotes the formation of hexatic phase \cite{lee2008effect}. Besides particle interaction and temperature, density and pressure affect the melting scenario. For example, soft-core particles can form multiple crystal phases that melt via the KTHNY scenario at high densities and exhibit discontinuous hexatic-liquid transition at low densities; \cite{zu2016density} The hexatic regime shrinks and vanishes as the temperature increases. \cite{zu2016density} 

  Besides the above studies on isotropic particles, particle's anisotropy also affects the 2D melting scenario. Anisotropic interactions (such as ionic, covalent, or hydrogen bonds) and multibody interactions are prevalent in atoms and molecules. A simple type of anisotropic interaction is induced by non-spherical particle shape. Simulations show that monodispersed hard regular polygons exhibit three melting scenarios: continuous solid-hexatic and first-order hexatic-fluid transitions for polygons with $\ge$7 edges; continuous solid-hexatic/tetratic/hexatic and hexatic/tetratic/hexatic-liquid transitions for hexagons/squares/triangles respectively; one-step first-order melting for pentagons. \cite{anderson2017shape} A tetratic phase has 4-fold symmetry and similar translational and orientational orders to those of a hexatic phase. 
  
  A recent Monte Carlo simulation published in \textit{Cell Reports Physical Science} 4, 10  (2023) \cite{jiang2023five} studied monodispersed hard truncated rhombs with different truncation parameters and observed two crystalline phases, two hexatic phases, and smectic and nematic phases (Fig.~1B). Note that anisotropic particles possess an additional degree of freedom about particle orientation, which can produce rotator phases, plastic crystal, nematic, smectic, and other liquid-crystal phases. The quasi-long-range order in either particle orientation or bond orientation between center of neighboring particles correspond to different hexatic phases. These particles exhibit five 2D melting scenarios including three known types (the first-order one-step transition; KTHNY scenario; and a continuous solid-to-metaphase and first-order
metaphase-to-liquid transition) and two novel types (a first-order solid-mesophase transition followed with either a continuous or first-order mesophase-isotropic fluid transition). The phases and melting behaviors are dominated by the positional 
entropy for rounded particles and by the orientational entropy for elongated particles. These interesting discoveries highlight the diverse 2D melting behaviors and are helpful for the future development of 2D melting theories. The results can be experimentally tested by future experiments using colloidal polygons printed by photolithography. Besides these studies about anisotropic particle shapes, how anisotropic or multibody interactions affect 2D melting deserves future studies. For particles with complicated interactions or shapes, various mesophases may exist between crystal and liquid, e.g. see Fig.~1. Consequently, various melting scenarios with multiple intermediate phases can exist (Fig.~1A). Strictly speaking, such ``melting scenario" contains multiple transitions instead of a crystal melting transition.

    \begin{figure*}[!t] 
		\centering
		\includegraphics[width=2.0 \columnwidth,]{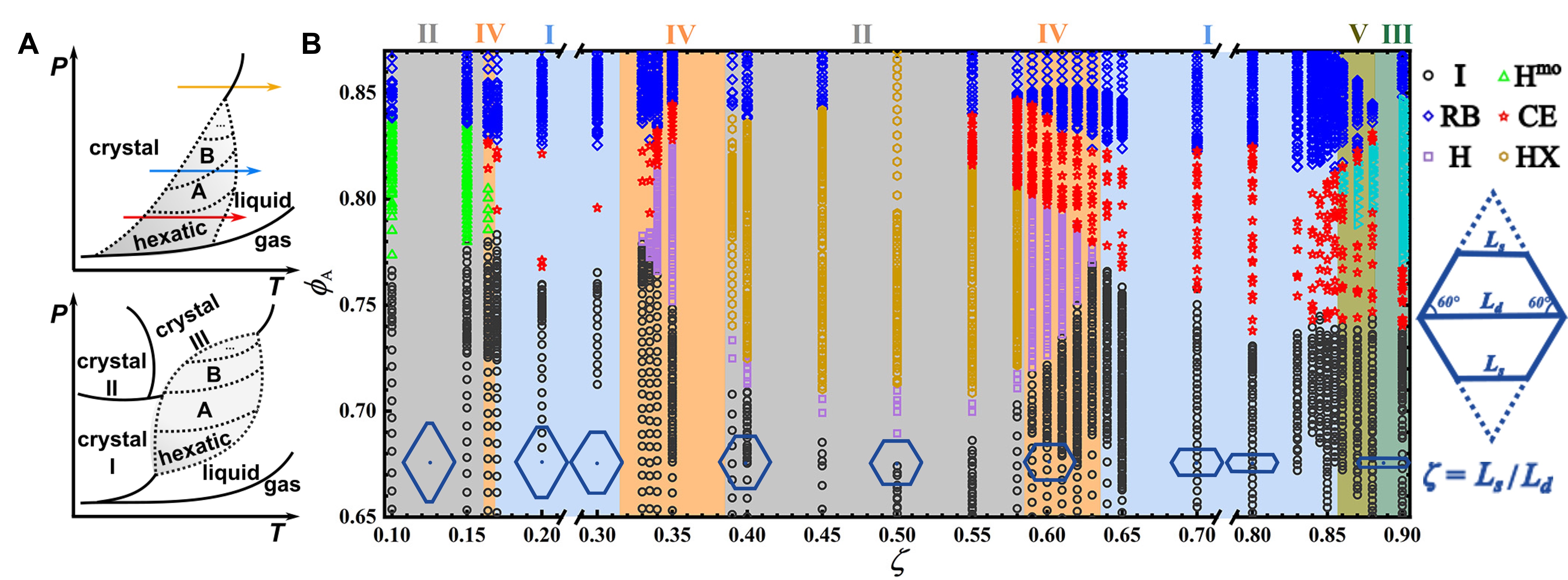}
		\caption{(A) Sketches of two possible phase diagrams. Many crystalline phases usually exist at high pressures. Increasing temperature or decreasing pressure often reduces the order in some degrees of freedom, resulting in partially disordered mesophases (labelled by hexatic, A, B...) between crystal and liquid. Thus, a crystal may need to pass some mesophases before transforming into a liquid, and these transitions (dotted curves) can be first-order or continuous. For a first-order transition, there is a two-phase coexistence regime. The KTHNY, one-step and other mesophase-mediated melting scenarios are labelled with red, yellow, and blue arrows, respectively. (B) Phase diagram of more than 50 different shapes of truncated rhombs.\cite{jiang2023five} For each type of monodispersed particles featured with a shape parameter $\xi$, decreasing the area fraction $\phi$ induces a 2D melting. I: isotropic fluid, $\rm H^{mo}$: hexatic phase in molecular-orientational order, RB: rhombic crystal, CE: coexistence of two neighboring phases, H: hexatic phase, HX: hexagonal crystal, Sm/N: smectic or nematic phase. These polygons exhibit five melting scenarios, including two novel types: (1) a first-order RB to H/$\rm H^{mo}$ and a continuous H/$\rm H^{mo}$ to I transition; (2) first-order RB to Sm/N and Sm/N to I transitions. }   
		\label{fig1}
	\end{figure*}   

    Weak first-order and continuous transitions often exist in 2D melting because of long-wavelength fluctuations, and it is challenging to distinguish them. Thus, a large system or sensitive indicator is often needed for conclusive results. There are various analysis methods for identifying the hexatic phase and the transition order, such as the spatial and time correlations of order parameters, elastic moduli, the peak shape of the structure factor, the density fluctuations and the two-phase coexistence. In particular, the Mayer-Wood loop in the equation of state can identify the order of transition relatively accurately.\cite{kapfer2015two} For example, it has been used to resolve the weak first-order hexatic-liquid transition in hard disks whose hexatic-liquid interfaces are very rough and blurry in the two-phase coexistence. 
    
    2D melting has also been explored in curved spaces, on solid substrates, with randomly pinned particles, and compared with  multilayer thin-film melting.\cite{wang2016melting} Some important aspects of 2D melting remain poorly, for instance, the kinetic pathway, heterogeneous melting due to defects and crystal surfaces, premelting\cite{li2016modes}, and non-equilibrium melting transitions under external fields or in active-particle systems. Surface premelting has been experimentally observed at the single-particle level for the first time by using colloids~\cite{li2016modes}, which revealed that monolayer and multilayer crystals have distinct behaviors in both surface premelting and bulk melting. This finding was subsequently confirmed by a simulation.

   2D crystal melting has of theoretical interest and practical importance in understanding low-dimensional materials. It has been studied in monolayers of atoms, molecules, electrons, superconducting vertices, skyrmions, plasma, colloids, and granular particles. However, new melting scenarios could still be discovered and some poorly explored aspects of 2D melting should be studied comprehensively in the future.

 ACKNOWLEDGMENTS
This work was supported by the Hong Kong Research Grants Council (C6016-20G).

DECLARATION OF INTERESTS
The authors declare no competing
interests.


    
\end{document}